\documentclass[twocolumn,amsmath,amssymb,superscriptaddress,floatfix, longbibliography]{revtex4-2}
\usepackage{printlen}
\usepackage[T1]{fontenc}
\usepackage[english]{babel}
\usepackage{graphicx}
\usepackage{geometry}
\usepackage{amsmath,amssymb,mathrsfs}
\usepackage[dvipsnames]{xcolor}
\usepackage[caption=false]{subfig}
\usepackage{float}
\usepackage{numprint}
\usepackage{siunitx}
\usepackage{todonotes}
\usetikzlibrary{matrix}
\usepackage{hyperref}
\usepackage{xr-hyper}   
\usepackage{soul}
\usepackage{array}
\usepackage{booktabs}
\usepackage[referable]{threeparttablex}
\usepackage{adjustbox}
\usepackage{tabularx}
\usepackage{multirow}
\usepackage{makecell}

\usepackage[version=4]{mhchem}

\usepackage{siunitx}

\makeatletter
\def\frontmatter@abstractheading{%
  \begin{center}\bfseries\MakeUppercase{\abstractname}\end{center}%
}
\makeatother

\begin{document}
\widetext

\title{First-order polarization process as an alternative to antiferroelectricity}

\author{{ Louis Bastogne}}
\thanks{These authors contributed equally to this work.}
\affiliation{Theoretical Materials Physics, Q-MAT, Universit\'e de Li\`ege, B-4000 Sart-Tilman, Belgium}%

\author{{Lukas Korosec }}
\thanks{These authors contributed equally to this work.}
\affiliation{Department of Quantum Matter Physics, University of Geneva, 24 Quai Ernest-Ansermet, 1211 Geneva, Switzerland}

\author{Yuchuan Wang}
\affiliation{Department of Physics, University of Warwick, Coventry, CV4 7AL, United Kingdom}

\author{Evgenios Stylianidis}
\affiliation{London Centre for Nanotechnology and Department of Physics and Astronomy, University College London, Gower Street, London WC1E 6BT, United Kingdom}
\affiliation{Solid State Physics Institute, TU Wien, Wiedner Hauptstr. 8-10/138, Vienna 1040, Austria}

\author{Daniel G. Porter}
\affiliation{Diamond Light Source, Harwell Science and Innovation Campus, Didcot, Oxfordshire OX11 0DE, United Kingdom}

\author{Gareth Nisbet}
\affiliation{Diamond Light Source, Harwell Science and Innovation Campus, Didcot, Oxfordshire OX11 0DE, United Kingdom}

\author{Cl\'ementine Thibault}
\affiliation{Department of Quantum Matter Physics, University of Geneva, 24 Quai Ernest-Ansermet, 1211 Geneva, Switzerland}

\author{{ Jean-Marc Triscone}}
\email{Jean-Marc.Triscone@unige.ch}
\affiliation{Department of Quantum Matter Physics, University of Geneva, 24 Quai Ernest-Ansermet, 1211 Geneva, Switzerland}%

\author{Marios Hadjimichael}
\email{marios.hadjimichael@warwick.ac.uk}
\affiliation{Department of Quantum Matter Physics, University of Geneva, 24 Quai Ernest-Ansermet, 1211 Geneva, Switzerland}
\affiliation{Department of Physics, University of Warwick, Coventry, CV4 7AL, United Kingdom}

\author{{ Philippe Ghosez }}
\email{Philippe.Ghosez@uliege.be}
\affiliation{Theoretical Materials Physics, Q-MAT, Universit\'e de Li\`ege, B-4000 Sart-Tilman, Belgium}%

\date{\today}

\begin{abstract}
Antiferroelectrics generate significant interest since their polarization versus electric field ($\mathcal{P}$-$\mathcal{E}$) curves show typical double-hysteresis loops appealing for various applications. Unfortunately, antiferroelectrics are rare.  In magnetic compounds, magnetization versus magnetic field ($\mathcal{M}$-$\mathcal{H}$) curves can show analogous double hysteresis loops not only in antiferromagnets but also in systems exhibiting field-induced first-order reorientation of the magnetization through a so-called first-order magnetization process.  Here, we show that appealing double-hysteresis $\mathcal{P}$-$\mathcal{E}$ loops can also appear from an  unprecedented first-order polarization process. 
Focusing on non-polar CaTiO$_3$, which can be turned ferroelectric under tensile strain, we study epitaxial thin films on differently-oriented NdGaO$_\mathrm{3}$ substrates using a combination of theoretical and experimental techniques. We uncover that a certain configuration exhibits double-hysteresis $\mathcal{P}$-$\mathcal{E}$ loops that we rationalize from a field-induced abrupt rotation of the polarization.
Such a first-order polarization process establishes a promising alternative pathway to achieve double hysteresis $\mathcal{P}$-$\mathcal{E}$ loops appealing for practical applications.
\end{abstract}

\maketitle

\section*{Introduction}
The concept of antiferroelectricity was introduced by Kittel in 1951~\cite{kittel1951theory} as the electric analogue of antiferromagnetism, and soon after exemplified in PbZrO$_\mathrm{3}$~\cite{shirane1951dielectric,sawaguchi1951antiferroelectric}.
Following the original idea of Kittel, antiferroelectrics are sometimes viewed as crystals with two sublattices with equal and opposite polarization, but not all such antipolar crystals are antiferroelectrics. 
According to classical textbooks, antiferroelectrics are more precisely defined as {\it antipolar systems that exhibit large dielectric anomalies near the Curie temperature and which can be transformed to an induced ferroelectric phase by application of an electric field} ~\cite{jona1962ferroelectric} or similarly, as {\it antipolar crystals whose free energy is closely comparable to that of a ferroelectric modification of the same crystal and which may therefore be switched from antipolar to ferroelectric by the application of an external electric field}~\cite{lines2001principles}.
Traditionally, their identification relies on observing double hysteresis loops in the polarization versus electric field ($\mathcal{P}$–$\mathcal{E}$) curves, a feature that appears nowadays particularly appealing for energy storage, electrocaloric applications, neuromorphic computing or water purification via piezocatalysis~\cite{zhuo2021perspective,xu2024novel,mischenko2006giant,zhang2025ferroelectric,amdouni2024enhancement}.

While antiferromagnets are quite generic among magnetic compounds, antiferroelectrics remain an exception among polar systems. In view of their timely interest for technological applications, different routes are currently explored for discovering new antiferroelectrics, including strategies to design artificial antiferroelectrics. This includes for instance, antipolar–polar transitions in superlattices~\cite{sigman2002antiferroelectric,aramberri2022ferroelectric,bousquet2010first,yin2024mimicking} or interface-induced effects~\cite{catalan2026modern,wu2015double}. In these examples, the strategy is often to tune the depolarizing field in nanostructures to slightly destabilize the polar state with respect to the antipolar one, while keeping the former accessible under finite field.

In magnetic systems, double-hysteresis loops  in magnetization versus magnetic field ($\mathcal{M}$-$\mathcal{H}$) curves are not restricted to antiferromagnets but can also appear as a feature of some ferromagnetic systems showing first-order magnetization process~\cite{asti1981some,asti1980theory}. In such systems, the free energy shows two slightly inequivalent extrema associated with distinct orientations of the magnetization. In some cases, the application of an appropriate external magnetic field can reverse the stability of the two inequivalent minima, producing a first-order transition from one state to the other through an abrupt rotation of the magnetization. In the case where this field-induced transition is reversible under removal of the magnetic field, it can give rise to double-hysteresis $\mathcal{M}$-$\mathcal{H}$ loops. %

One may wonder whether, by analogy, a similar type of so-called first-order polarization process could exist in polar materials, associated to a field-induced abrupt first-order rotation of the polarization and able to generate double hysteresis loops without invoking a conventional antipolar ground state. While a few early studies hinted at such possibilities in liquid crystals, amorphous systems or models inspired by biological systems~\cite{khan1996planar,chieu1988effect,tuszynski1985generalization}, this concept has remained largely unexplored in crystalline ferroelectrics. Demonstrating an electric analogue of the first-order magnetization process would not only broaden the fundamental understanding of polarization switching but also open new opportunities for designing materials with controllable hysteresis and enhanced functional responses.

To explore this idea, we focus on CaTiO$_\mathrm{3}$, the archetypal perovskite oxide, which exhibits a non-polar orthorhombic $Pbnm$ ground state involving oxygen octahedra rotations but is sometimes seen as an incipient ferroelectric~\cite{lemanov1999perovskite,zhang2025finite}. Lying close to a ferroelectric instability, it was shown to develop a spontaneous polarization under tensile strain or chemical substitution~\cite{eklund2009strain,gu2012phenomenological,PhysRevMaterials.8.094412,sarantopoulos2018effect,biegalski2015,lemanov2002incipient}. Considering epitaxial CaTiO$_\mathrm{3}$ films grown on differently oriented NdGaO$_\mathrm{3}$ orthorhombic substrates, we show, through a combination of experimental and theoretical approaches, that epitaxial strain can tune the energy and polarization landscapes, stabilize different oxygen octahedral rotation patterns, and ultimately enable deterministic first-order polarization process with associated appealing double hysteresis loops.

\section{results}

We first examine how tensile epitaxial strain can induce and stabilize different polar states in CaTiO$_\mathrm{3}$. In bulk, CaTiO$_\mathrm{3}$ adopts an orthorhombic $Pbnm$ phase which can be seen as a small distortion of the reference cubic perovskite structure: it arises from the condensation of in-phase and anti-phase rotations of the oxygen octahedra according to a pattern ($a^-a^-b^+$) in Glazer notation ~\cite{glazer1972classification}, and further includes secondary antipolar motions of the Ca-cations~\cite{miao2013first} in the plane perpendicular to the in-phase rotation axis of the oxygen octahedra.%

Previous theoretical~\cite{eklund2009strain,gu2012phenomenological,PhysRevMaterials.8.094412} and experimental~\cite{biegalski2015,haislmaier2016} studies have shown that, under epitaxial tensile strain, CaTiO$_\mathrm{3}$ preserves its oxygen octahedra rotation pattern but further develops polarization along different directions and exhibits competing ferroelectric phases. 
This testifies to a complex and tunable energy landscape and provides an appealing opportunity to explore the possibility of achieving a first-order polarization process.

\subsection{Epitaxial mechanical constraints}
\begin{figure*}[h!]
    \centering
    \includegraphics[width=\linewidth]{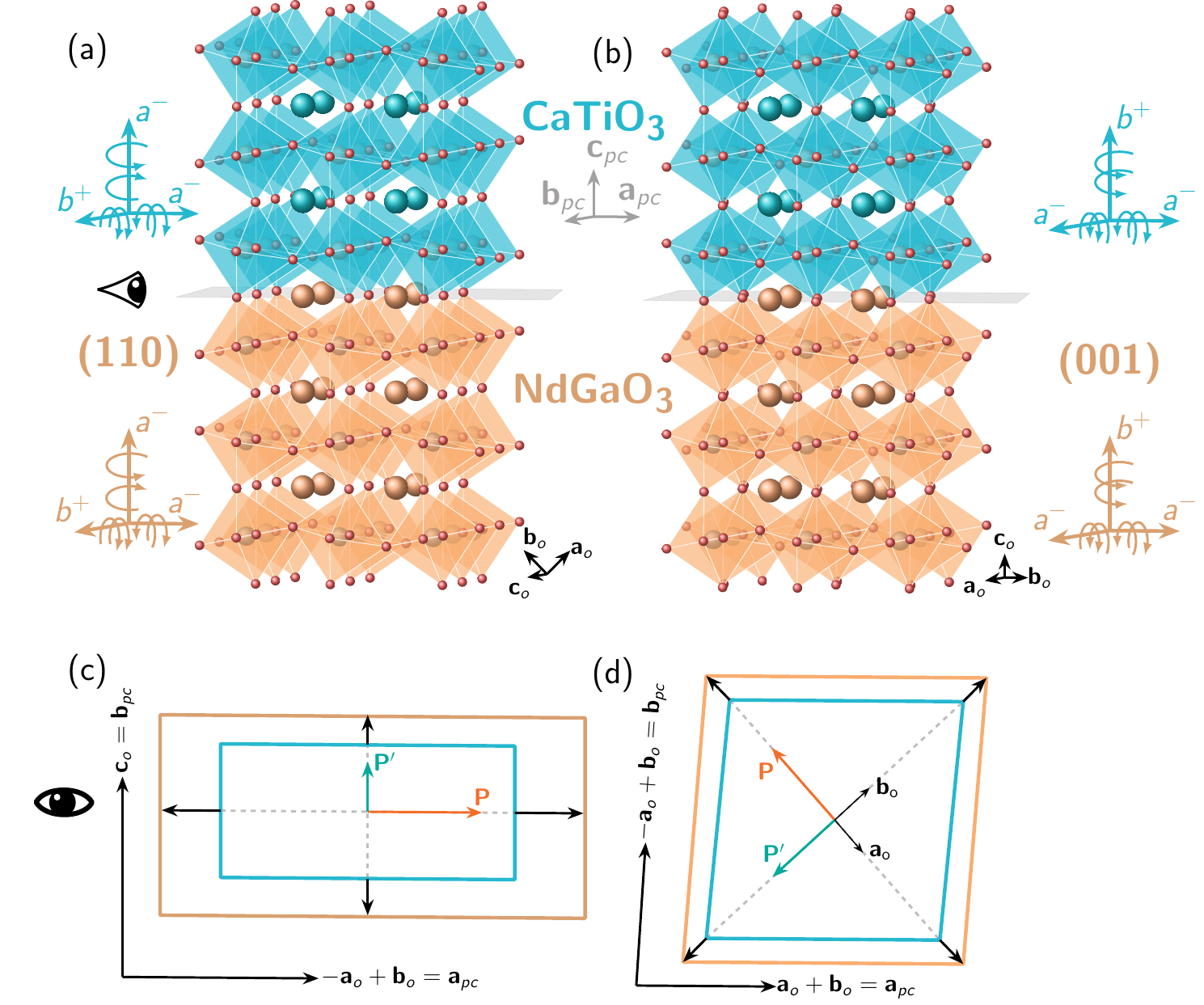}
    \caption{\textbf{Structure and epitaxial strain of CaTiO$_3$ films on
differently oriented NdGaO$_3$.} Sketch of atomic configuration and oxygen
octahedra orientation of CaTiO$_3$ films grown on (a) (110)$_{\text{o}}$- and
(b) (001)$_{\text{o}}$-oriented NdGaO$_3$. Tensile strain imposed by
(c) (110)$_{\text{o}}$- and (d) (001)$_{\text{o}}$-oriented NdGaO$_3$ as well
as the resulting dominant polarization orientations. $\mathbf{P}$ and
$\mathbf{P}'$ signify the primary and secondary polarization directions as
found experimentally and theoretically. The subscripts o and pc denote
orthorhombic and pseudocubic respectively. The eye symbol indicates the
viewing direction used.}
    \label{fig:NGO_110_and_NGO_001_config}
\end{figure*}

To investigate this phenomenon, we have deposited high-quality CaTiO$_\mathrm{3}$ thin films on (110)$_\text{o}$ and (001)$_\text{o}$-oriented NdGaO$_\mathrm{3}$ substrates, with high structural quality and atomically flat surfaces (see SI, Figs.~S1 and S2). The NdGaO$_\mathrm{3}$ substrate adopts the same orthorhombic $Pbnm$ phase as bulk CaTiO$_\mathrm{3}$. Simple inspection of the lattice parameters (see SI, Table~S1) suggests that, when epitaxially grown on (110)$_{\text{o}}$- and (001)$_{\text{o}}$-oriented  NdGaO$_3$, CaTiO$_3$ should keep in each case the same orientation as the substrate to minimize elastic energy. 
Our second-principles calculations support this expectation and show that by imposing only epitaxial strain, the most favorable orthorhombic unit cell orientation of the film is the one that follows the orientation of the substrate (see SI, Table~S2). 
In practice, this additionally preserves the continuity of the pattern of oxygen octahedra rotations across the interface (Fig. ~\ref{fig:NGO_110_and_NGO_001_config}), which should further stabilize the orientation. For both orientations, X-ray diffraction (XRD) confirms that the films are fully strained, twin-free, and inherit the substrate octahedral rotation pattern across the interface (see SI, Figs.~S3--S5).

Importantly, although they arise from the same material, the mechanical constraints imposed on the CaTiO$_3$ film  by the NdGaO$_3$ substrate are significantly different for the two substrate orientations (see SI, Table~S2) and should therefore affect the energy landscape differently. The (110)$_{\text{o}}$ substrate applies a rectangular biaxial tensile strain along $\mathbf{c}_{\text{o}}$ and $\mathbf{b}_{\text{o}}-\mathbf{a}_{\text{o}}$, while the (001)$_{\text{o}}$ substrate produces an orthorhombic tensile constraint in the ($a_{\text{o}},b_{\text{o}}$) plane. This difference in the degree of strain leads to markedly different behavior for the two orientations of films, as will be shown below.

\subsection{CaTiO$_\mathrm{3}$ on (110)$_\text{o}$-oriented NdGaO$_\mathrm{3}$} 

We first focus on CaTiO$_\mathrm{3}$ films grown on (110)$_\mathrm{o}$-oriented NdGaO$_\mathrm{3}$ (see Methods).  Combined XRD (see SI, Fig.~S3), electrical measurements (see below), and first-principles calculations demonstrate that this strain state stabilizes a low-symmetry $Pm$ polar phase with an in-plane polarization along $\mathbf{a}_\mathrm{pc}$ [Fig.~\ref{fig:NGO_110_and_NGO_001_config}(b)], slightly lower in energy than a competing $P2_1$ phase polarized along $\mathbf{b}_\mathrm{pc}$.

\begin{figure*}[h!]
    \centering
    \includegraphics[width=\textwidth]{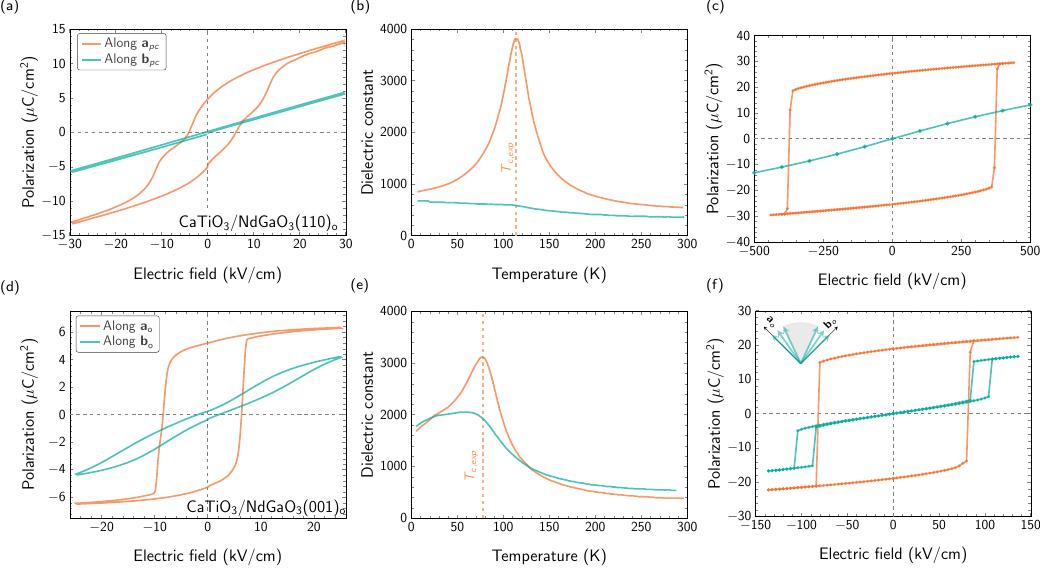}
    \caption{\textbf{Characterisation of the ferroelectric properties of CaTiO$_3$ films.}
    (a) Experimental polarization versus electric field measurement for a 18.2 nm-thick CaTiO$_\mathrm{3}$ film  grown on (110)$_{\text{o}}$-oriented NdGaO$_\mathrm{3}$ along $\mathbf{a}_{\text{pc}}$ and $\mathbf{b}_{\text{pc}}$ at $T_{\mathrm{exp}}$ = 4.2 K in $Pm$ phase. 
    (b) Experimental dielectric constant versus temperature measurement for the same film as panel (a), along $\mathbf{a}_{\text{pc}}$ and $\mathbf{b}_{\text{pc}}$, showing a characteristic peak at $T_{c,exp}$ = 115 K signifying the transition from $Pm$ to $P2_1/m$ phase. 
    (c) Second-principles polarization versus electric field curve for CaTiO$_\mathrm{3}$ strained on (110)$_{\text{o}}$-oriented NdGaO$_\mathrm{3}$ along $\mathbf{a}_{\text{pc}}$ and $\mathbf{b}_{\text{pc}}$ at $T_{\mathrm{sim}}$ = 40~K in $Pm$ phase.  
    (d) Experimental polarization versus electric field measurement for a 17.4 nm-thick CaTiO$_\mathrm{3}$ film deposited on (001)$_{\text{o}}$-oriented NdGaO$_\mathrm{3}$  along $\mathbf{a}_{\text{o}}$ and $\mathbf{b}_{\text{o}}$ at $T_{exp}$ = 4.2 K in $Pm$ phase. Note the different horizontal axis scale compared to panel (a). 
    (e) Experimental dielectric constant versus temperature measurement for the same film as panel (d), along $\mathbf{a}_{\text{o}}$ and $\mathbf{b}_{\text{o}}$ showing a characteristic peak at $T_{\mathrm{c,exp}}$ = 80 K signifying the transition from $Pm$($a_o$) to $Pbnm$ phase. 
    (f) Second-principles polarization versus electric field curve for CaTiO$_\mathrm{3}$ strained on NdGaO$_\mathrm{3}$ (001)$_{\text{o}}$ along $\mathbf{a}_{\text{o}}$ and $\mathbf{b}_{\text{o}}$ at $T_\mathrm{{sim}}$ = 40~K in $Pm$($a_o$) phase. Note the different horizontal axis scale compared to panel (c). Orange and teal curves correspond to fields applied along the first and second
in-plane direction indicated in each row, respectively.}
    \label{fig:NGO_001_elec_and_thermal_mes}
\end{figure*}

Having resolved the structure, we next examine the electrical response of an 18.2 nm thick film along the two orthogonal in-plane axes. Fig.~\ref{fig:NGO_001_elec_and_thermal_mes}(a) shows that a ferroelectric hysteresis loop is obtained at 4.2 K when the field is applied along $\mathbf{a}_{\text{pc}}$, whereas only a linear dielectric response is observed along $\mathbf{b}_{\text{pc}}$. Capacitance–voltage measurements (SI, Fig.~S9) and second-principles simulations [Fig.~\ref{fig:NGO_001_elec_and_thermal_mes}(c)] further confirm this anisotropic response, demonstrating that CaTiO$_\mathrm{3}$ behaves as a uniaxial ferroelectric under this strain state.

We note that the ferroelectric hysteresis loop exhibits a step-like or partially pinched shape [Fig.~\ref{fig:NGO_001_elec_and_thermal_mes}(a)]. Such pinched loops have previously been reported in ferroelectric thin films and ceramics and are often associated with extrinsic effects, such as defect dipoles or inhomogeneous switching conditions~\cite{carl1977electrical,morozov2008hardening,everhardt2020temperature}. Through further frequency-dependent switching and fatigue measurements we have found that the degree of pinching in our hysteresis loops varies as a function of frequency and sample switching history (SI, Fig.~S10). This suggests that these pinched hysteresis loops are not intrinsic to CaTiO$_\mathrm{3}$ under this strain state but are most likely associated with extrinsic effects~\cite{morozov2008hardening}.

The temperature dependence of the dielectric constant along the two orthogonal directions shown in Fig.~\ref{fig:NGO_001_elec_and_thermal_mes}(b) further supports the uniaxial ferroelectric behavior: a pronounced peak appears only when the field is applied along the ferroelectric axis $\mathbf{a}_{\text{pc}}$, while the curve remains flat when the field is applied along the orthogonal $\mathbf{b}_{\text{pc}}$ direction (see SI, Fig.~S11 for frequency-dependent capacitance and loss tangent versus temperature, showing strong anisotropic response along the two orthogonal directions). This further identifies the ferroelectric Curie temperature to $T_{c,exp} = 115$ K, a value slightly overestimated by our second-principles model ($T_{c,sim}= 245$ K -- see SI, Fig. ~S12). Together, these results highlight the strong directional coupling between strain, octahedra rotations, and polarization in such orientation.

\subsection{CaTiO$_\mathrm{3}$ on (001)$_\text{o}$-oriented NdGaO$_3$} 

We now turn to CaTiO$_\mathrm{3}$ films epitaxially strained on (001)$_\mathrm{o}$-oriented NdGaO$_\mathrm{3}$ (SI, Fig.~S5). Below approximately 87 K, synchrotron XRD measurements reveal a structural phase transition that lowers the symmetry to a polar $Pm$ phase  (SI, Fig.~S6 and~S7), in excellent agreement with second-principles calculations which yield a transition temperature of 85~K. The polarization is predominantly oriented along $\mathbf{a}_\mathrm{o}$  and shows a smaller component along $\mathbf{b}_\mathrm{o}$. We note that synchrotron XRD reciprocal space maps around thin film Bragg peaks at 15 K do not show evidence of diffuse scattering as expected in case of domain formation in the thin films.

First- and second-principles calculations corroborate this picture and further reveal the presence of two nearly degenerate polar $Pm$ phases, separated by only 0.7 meV/f.u. [Fig.~\ref{fig:NGO_110_and_NGO_001_config}(d)]. One is polarized primarily along $a_\mathrm{o}$ (close to $P2_1nm$), the other along $b_\mathrm{o}$ (close to $Pb2_1m$). Their near degeneracy arises from a subtle competition: in the $P2_1nm$-like state, the main polarization component is perpendicular to the antipolar motions of the Ca cations, whereas it is parallel to them in the $Pb2_1m$-like state. In what follows, in order to avoid any ambiguity between both $Pm$ phases, we refer to them respectively as $Pm$($a_o$) and $Pm$($b_o$), according to the direction of the main component of their polarization.

We next examine the electrical and dielectric measurements which reveal a marked anisotropy: while a conventional ferroelectric loop is observed when the field is applied along $\mathbf{a}_\mathrm{o}$, a double hysteresis loop emerges when the field is applied along $\mathbf{b}_\mathrm{o}$ [Fig.~\ref{fig:NGO_001_elec_and_thermal_mes}(d–e)]. This distinct behavior is consistent with capacitance–voltage measurements (SI, Fig.~S13) and is reproduced in second-principles simulations [Fig. \ref{fig:NGO_001_elec_and_thermal_mes}(f)],
which reproduce a simple ferroelectric loop along $\mathbf{a}_\mathrm{o}$ and the emergence of a double hysteresis loop when the electric field is applied along $\mathbf{b}_{\text{o}}$. Importantly, unlike the case of CaTiO$_\mathrm{3}$ grown on (110)$_\mathrm{o}$-oriented NdGaO$_\mathrm{3}$, the ferroelectric and antiferroelectric hysteresis loops along $\mathbf{a}_\mathrm{o}$ and $\mathbf{b}_\mathrm{o}$ do not change shape or magnitude with frequency and field cycling, with switching behaviour that remains stable up to $10^7$ cycles (SI, Fig.~S14 and S15). This indicates that the behavior that is observed in these films is intrinsic and does not arise from any extrinsic effects such as aging or polarization pinning defects~\cite{morozov2008hardening}.
Temperature-dependent dielectric measurements in Fig. \ref{fig:NGO_001_elec_and_thermal_mes}(e) provide an independent manifestation of this duality and anisotropic response, revealing markedly different dielectric signatures depending on the orientation of the applied electric field. This anisotropy is additionally evident from the frequency dependent capacitance and loss tangent (see SI, Fig.~S16). Furthermore, temperature-dependent switching experiments show that both directions become paraelectric above approximately 85 K, consistent with permittivity and synchrotron measurements (see SI, Figs. S6, S17 and S18).

The pronounced directional switching observed here is unusual. This behavior evokes features typical of first-order magnetization process, but alternative mechanisms could, in principle, produce such a behavior. In particular, Ref.~\cite{yin2024mimicking} shows that, in ferroelectric/dielectric superlattices, competition between domain configurations and field-induced reorientation, governed by electrostatic and mechanical boundary conditions, can lead alternatively to single or double hysteresis loops depending on the orientation of the applied electric field.

\begin{figure*}[h!]
  \centering
    \includegraphics[width=\textwidth]{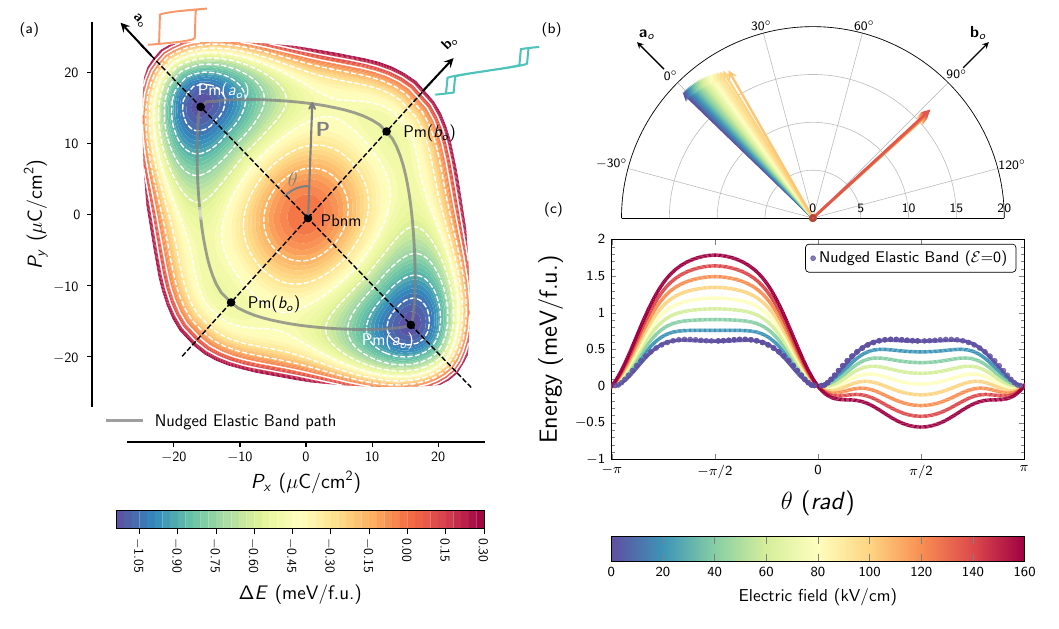}
     \caption{\textbf{Polarization energy landscape and first-order
    polarization process in CaTiO$_3$ strained on
    (001)$_\mathrm{o}$-oriented NdGaO$_3$.} (a) Energy landscape of CaTiO$_\mathrm{3}$ strained on NdGaO$_\mathrm{3}$ (001)$_\mathrm{o}$, shown as a function of the pseudo-cubic polarization components $P_\text{x}$ and $P_\text{y}$. The grey line marks the lowest-energy path (nudged elastic band) connecting the $Pm$($a_o$) ground state and the $Pm$($b_o$) metastable phase, defining a polar angle $\theta$ that parameterizes the polarization rotation along this path. The landscape exhibits a Mexican-hat-like shape (although not perfectly flat). For electric fields applied along $\mathbf{a}_\mathrm{o}$ (resp. $\mathbf{b}_\mathrm{o}$), the system exhibits single (resp. double) hysteresis loops.
(b) Reorientation of the polarization in CaTiO$_\mathrm{3}$ strained on NdGaO$_\mathrm{3}$ (001)$_\mathrm{o}$ under electric field along $\mathbf{b}_\mathrm{o}$ at $T_{\mathrm{sim}}=40$ K [Fig.~\ref{fig:NGO_001_elec_and_thermal_mes}~(f)], highlighting an abrupt jump associated with a first-order transition.
(c) Energy along the same minimum-energy path shown in panel (a), plotted as a function of the rotation angle $\theta$, and its evolution under electric field applied along $\mathbf{b}_\mathrm{o}$, as fitted with Eq.~S1.}
    \label{fig:FOPP_summary}
\end{figure*}

To assess the origin of this behavior in CaTiO$_\mathrm{3}$ films, we perform a fine sampling of the polarization energy landscape [Fig.~\ref{fig:FOPP_summary} (a)] and look for the associated switching pathways. We first compute the minimum-energy path connecting the two experimentally accessible polarization states ($Pm$($a_o$) ground state and $Pm$($b_o$) metastable phase) using second-principles nudged elastic band (NEB) calculations [Fig.~\ref{fig:FOPP_summary} (a,c)]. The resulting path reveals a Mexican hat-shape energy landscape, which is compatible with a progressive in-plane rotation of the polarization vector between the nearly orthogonal minima, without passing through the paraelectric $Pbnm$ state, when a field is applied along $\mathbf{b}_\mathrm{o}$.
This polarization rotation is further confirmed by field-driven molecular-dynamics simulations. Finite-temperature simulations under electric field applied along $\mathbf{b}_\mathrm{o}$ [Fig.~\ref{fig:NGO_001_elec_and_thermal_mes}(f)] show that the polarization rotates progressively from $\mathbf{a}_\mathrm{o}$ to $\mathbf{b}_\mathrm{o}$ with increasing field until a critical field is reached, where it abruptly jumps from the $Pm$($a_o$) state to the $Pm$($b_o$) state, while remaining finite throughout the transition and never passing through zero. The polarization trajectories shown in Fig.~\ref{fig:FOPP_summary}(b) provide an alternative representation of the same data, directly visualizing this continuous and almost rigid rotation followed by the sudden jump.
Together, these results establish that the observed double hysteresis loop is associated with a discontinuous reorientation of the polarization between two ferroelectric states of distinct orientations that are close in energy and separated by a small energy barrier. By analogy with magnetism, we propose to call this phenomenon a {\it first-order polarization process}.

\section{Discussion}

Our experimental and theoretical results consistently establish that CaTiO$_\mathrm{3}$ films grown on NdGaO$_\mathrm{3}$ (001)$_{\text{o}}$ substrates exhibit either conventional ferroelectric switching or double hysteresis loops, depending on the orientation of the applied electric field. This directional duality arises from a  first-order polarization process, an analogue of first-order magnetization process in magnetic systems. 
This interpretation is supported by a one-to-one correspondence between the key characteristics of first-order magnetization process and our observations in CaTiO$_\mathrm{3}$.

The foundation of this analogy lies in an underlying energy landscape prone to polarization rotation. Our calculations reveal two nearly degenerate, yet non-equivalent, $Pm$ minima associated with nearly perpendicular polarization orientations and separated by a small energy barrier.

Extending the analogy to a more quantitative level, the polarization energy landscape can be described within a phenomenological framework using a mathematical formalism directly analogous to that employed for first-order magnetization process. To capture the NEB energy profile $U$ along the minimum-energy path [Fig.~\ref{fig:FOPP_summary} (a)] in a compact form, we consider that it only depends on the orientation $\theta$ -- with respect to $\mathbf{a}_\mathrm{o}$ -- of the spontaneous polarization of amplitude $P_s$, assumed to be constant, and we fit $U(\theta)$ using a minimal angular expansion:
\begin{equation}
U(\theta) = U_0 + K_1 \sin^2\theta + K_2 \sin^4\theta + K_3 \sin^6\theta
\label{eq:FOPP}
\end{equation}
This fit (Table~S3) perfectly reproduces both the relative depth of the two minima and the barrier separating them [Fig.~\ref{fig:FOPP_summary} (b)], providing a concise representation of the anisotropy governing polarization rotation.

Assuming a constant polarization amplitude $P_s$, the coupling to an external electric field ${\cal E}$ along $\mathbf{b}_\mathrm{o}$ is introduced through the field-dependent energy $F(\theta) = U(\theta) - {\cal E} P_s \sin\theta$. Within this phenomenological framework, the application of an external electric field progressively tilts the energy landscape, leading to a field-induced interchange of stability between the two competing minima at $\mathcal{E}\sim140$~kV/cm at 0~K, in good agreement with the coercive field obtained from second-principles molecular dynamics simulations. This agreement confirms that, at the phenomenological level, the double hysteresis loop is governed by an abrupt and reversible reorientation of the polarization vector that is directly analogous to the magnetization reversal characteristic of a first-order magnetization process (Fig.~S20). As usually observed~\cite{landauer1957electrostatic,scott2016review} and owing to the spatial and temporal limitations of the simulations, the theoretical coercive field overestimates the one observed experimentally, without affecting the physical interpretation or conclusions of our work.

This unified framework also explains the absence of double hysteresis loops in films grown on (110)$_\text{o}$ substrates. Second-principles NEB calculations reveal a substantially higher energy barrier between the competing polarization states ($P2_1$ and $P_m$), approximately four times larger than in the (001)$_\text{o}$ case (see SI, Fig.~S19). Within the same parametrization, this increased barrier translates into a predicted coercive field exceeding 500~kV/cm at 0~K, well above the dielectric breakdown threshold of our CaTiO$_3$ thin films. As a result, the practically accessible electric fields cannot induce a transition toward the competing polarization states, and only uniaxial ferroelectric switching is experimentally observed in this orientation. Such in-plane uniaxial ferroelectric thin films offer a simplified polarization landscape and can be advantageous for applications requiring directional control, such as non-volatile memories, electro-optical modulators and in-plane tunnel junctions~\cite{buse1998non,shen2019plane,liu2023plane,wessels2007ferroelectric}.

These contrasting cases suggest concrete guidelines for realizing first-order polarization processes in ferroelectrics: (i) selecting materials with intrinsic structural anisotropies that differentiate polarization orientations, such as ferroelastic or antipolar distortions; (ii) using epitaxial boundary conditions to tune the relative stability of the resulting ferroelectric states toward near degeneracy without suppressing their polar character; and (iii) ensuring that the barriers separating these states remain moderate, so that polarization reorientation can be driven reversibly by experimentally accessible electric fields. From this perspective, a broad class of strained ferroelastic and antipolar materials emerges as promising candidates for first-order polarization processes~\cite{benedek2013there}.

CaTiO$_\mathrm{3}$ provides a concrete realization of these guidelines. The oxygen-octahedra rotation pattern, inherent to its orthorhombic structure, introduces a built-in anisotropy between different polarization directions, while epitaxial strain acts as an external and tunable control parameter that can selectively enhance or partially compensate this intrinsic anisotropy depending on its magnitude and symmetry~\cite{eklund2009strain}. The resulting balance stabilizes multiple ferroelectric states of comparable energy while maintaining moderate barriers between them, thereby enabling a first-order polarization reorientation under moderate applied electric fields.

To further assess the generality of our findings, we have explored theoretically SrTiO$_\text{3}$ on $(001)_\text{o}$-oriented DyScO$_\text{3}$ as an alternative candidate. The anisotropic tensile strain imposed by the substrate stabilizes two slightly inequivalent polar states with orthogonal in-plane polarizations.  Second-principles calculations predict an analogous polarization rotation and a double hysteresis loop, which weakens and eventually disappears as temperature increases (see Supplementary Fig.~S22). This confirms first-order polarization process as an alternative promising pathway to achieve double hysteresis loops.

Viewed from this perspective, the first-order polarization process identified here does not rely on an exotic switching pathway, but instead on a well-established mechanism in ferroelectrics, namely polarization rotation. Polarization rotation is indeed not unusual in ferroelectric perovskites and has been extensively studied in connection with giant piezoelectric responses~\cite{fu2000polarization,park1997ultrahigh,noheda2001polarization}. In those systems, the rotation of the polarization vector under an electric field is the key mechanism enabling ultrahigh electromechanical coupling, but it typically requires very large fields to overcome substantial intrinsic energy barriers~\cite{sai2002theory,ma2014electric}. In contrast, the strain-engineered energy landscape considered here brings competing ferroelectric minima into close energetic proximity, enabling a field-induced polarization reorientation at much lower, experimentally accessible fields. As a result, polarization rotation manifests not as a smooth response, but as a first-order switching process, giving rise to double hysteresis loops.

The discovery of a first-order polarization process immediately raises the question of its technological relevance, since the coexistence of distinct switching modes can be harnessed in practical functionalities.The relevant figures of merit extracted from our experimental and simulated hysteresis loops are summarized in Table~S4.
Importantly, the first-order polarization process is not only an alternative to antiferroelectricity.
The ability to toggle between ferroelectric-like and antiferroelectric-like responses depending of the orientation of the applied field within the same system is particularly appealing for neuromorphic computing, where ferroelectric switching underlies convolutional neural networks, while antiferroelectric-like behavior can emulate spiking neural networks—two complementary elements for hybrid neuromorphic architectures~\cite{zhang2025ferroelectric}.

Moreover, the duality of the dielectric response extends beyond polarization switching. Indeed, in magnetic systems, first-order magnetization process is often accompanied by magnetocaloric of opposite sign depending of the orientation of the applied field effects~\cite{ilyn2008magnetocaloric,terentev2015magnetocrystalline}; by analogy, first-order polarization process is expected to couple to similar electrocaloric behaviors. Consistent with this picture, we predict from our second-principles model (see methods) opposite signs of the electrocaloric temperature change depending on field orientation: $\Delta T < 0$  when the electric field is applied perpendicular to the polarization and is in the same range of efficiency compared to conventional antiferroelectics (see Table~S4 for a summary of the relevant figures of merit)~\cite{chen2021maxwell,li2022giant}. Conversely, our simulations predict $\Delta T > 0$ when the electric field is parallel to the polarization as shown in Fig.~S21(b) (see Table S4 for a summary of the relevant figures of merit). 
This mirrors the behavior of first-order magnetization process. It
further reinforces the close analogy between first-order magnetization process and first-order polarization process and highlights strained CaTiO$_\mathrm{3}$ as a promising platform for multifunctional applications.

This work demonstrates that antiferroelectric-like double hysteresis behavior can emerge in a purely ferroelectric material through a first-order polarization process, without invoking an antiferroelectric ground state. By strain-engineering the polarization energy landscape of CaTiO$_\mathrm{3}$, we show that multiple ferroelectric states of distinct orientation can be brought into close energetic competition, enabling field-driven switching between single and double hysteresis responses.
A key outcome is that ferroelectric and antiferroelectric-like functionalities coexist within the same material, under identical structural and thermodynamic conditions, and can be selectively accessed by the orientation of the applied electric field. This duality offers a potential platform combining energy storage, non-volatile memory operation, electrocaloric effects of both signs, and neuromorphic functionalities within a single ferroic system.
More broadly, our results establish first-order polarization processes as a general framework for designing multifunctional ferroelectrics, and suggest that strain-engineered ferroelastic and antipolar materials constitute a fertile playground for discovering analogous behavior beyond conventional antiferroelectrics.

\section{Methods}
\subsection{Thin film deposition and structural characterization}

Thin films of CaTiO$_\mathrm{3}$ were deposited in a home-built off-axis radio-frequency magnetron sputtering system, using \qty{50}{\watt} RF power and a 20:29~\ce{O2}:\ce{Ar} mixture at a pressure of 0.18 Torr, with a substrate temperature between 740 to 780~$^\circ$C. Commercial epi-polished substrates were used as-received, without further annealing or surface treatment.
NdGaO$_\mathrm{3}$~(110)$_\mathrm{o}$ substrates were purchased from SurfaceNet GmbH.
NdGaO$_\mathrm{3}$~(001)$_\mathrm{o}$ were purchased from Crystec GmbH.

High-resolution laboratory x-ray diffraction measurements were performed on a Panalytical X'Pert Pro diffractometer with a Cu-$K\alpha$ source. Atomic force microscopy topography measurements were performed using tapping mode on a Veeco Multimode IV AFM equipped with a SPECS Nanonis 4 controller using MikroMasch HQ:NSC15/Al BS AFM tips with a force constant of \qty{\approx 40}{\newton\per\meter}.

\subsection{Electrical measurements}

Platinum interdigitated electrodes with \numproduct{2 x 25} fingers with \qty{10}{\micro\metre} width, \qty{10}{\micro\metre} spacing, and \qty{1}{\milli\metre} length were fabricated by a standard photolithographic lift-off process.
Capacitance measurements as a function of temperature and applied voltage were performed using a Agilent E4980A LCR meter, applying a 1~V excitation at 1~kHz.
Polarization loops were measured using dynamical hysteresis measurements (DHM) at varying frequencies with an aixACCT TF2000 thin film ferroelectric tester. 
For both capacitance and polarization measurements, a home-built sample holder was used that can be inserted into a liquid helium dewar for cooling. Precise measurements of the temperature dependence of the remanent polarization were performed in an attoDRY closed cycle cryostat with sample temperature controlled using a Lakeshore 335 temperature controller, using an aixACCT TF2000 ferroelectric tester with triangular excitation voltage applied at varying frequencies. 

\subsection{Analysis of electrical measurements}

In order to determine dielectric constants, we use the Python implementation of the electrostatic model for interdigitated electrodes of Raeder et al.~\cite{reader2020, reader_python_2020}. Above the electrodes, we always simulate a semi-infinite layer of vacuum ($\epsilon = 1$).

 To determine the dielectric constant of the NdGaO$_\mathrm{3}$ (110)$_\mathrm{o}$ substrate, we simulate a semi-infinite layer below the electrodes and assume that the in-plane and out-of-plane dielectric constants are equal.
For measurements along the [$\bar{1}10$]$_\mathrm{o}$ axis, this assumption is justified because the in-plane and out-of-plane directions are equivalent by symmetry.
For measurements along the [001]$_\mathrm{o}$ axis, it is justified a posteriori by the fact that the resulting dielectric constants are the same along both measured directions within experimental accuracy below 115~K, where our CaTiO$_\mathrm{3}$ films become ferroelectric, and remain similar up to room temperature (see Fig.~S8 in the Supplementary Information.
We obtain the dielectric constant by numerically solving for the value that reproduces the measured capacitance at each temperature. For all analyses of CaTiO$_\mathrm{3}$ thin film samples, we treat the NdGaO$_\mathrm{3}$ substrates as linear isotropic dielectric media and use the experimental temperature-dependent dielectric constant measured along the [$\bar{1}10$]$_\mathrm{o}$ axis.

To determine the dielectric constant of CaTiO$_\mathrm{3}$ thin films from temperature- or voltage-dependent capacitance measurements, we simulate a finite layer with the thickness determined experimentally from X-ray diffraction on top of a semi-infinite substrate.
In a film with thickness ($\sim$20~nm) much smaller than the electrode spacing (10~$\mu$m), the electric field lies almost purely in plane~\cite{reader2020}.
Therefore, the out-of-plane dielectric constant of the film has a negligible effect on the total capacitance.
For simplicity, we thus set the out-of-plane and in-plane dielectric constants to be equal in the simulations.
Finally, we numerically solve for the dielectric constant of the thin film layer that reproduces the measured capacitance for each temperature or voltage.

To extract the polarisation-electric field loops of CaTiO$_\text{3}$ thin films
from the measured charge-voltage loops, we use two methods: (i) we
simulate the capacitance of an NdGaO$_\text{3}$ substrate using the
interdigitated-electrode electrostatic model described above and subtract
the corresponding charge from the experimental data, and (ii) we use the
conformal mapping method developed in Refs.~\cite{Veyres1980,
Schinzinger1991, Biegalski_Thesis_2006} and applied to interdigitated
electrodes in Ref.~\cite{zhang2024electrocaloric}. For method (ii),
starting from the expressions of capacitance in
Ref.~\cite{zhang2024electrocaloric}, using
\begin{equation}
P_{\mathrm{film}}=\varepsilon_0\left[\varepsilon_{\mathrm{film}}-1\right]E
\end{equation}

and 

\begin{equation}
C_{\mathrm{meas}}=\dfrac{Q_{\mathrm{meas}}}{V},
\end{equation}
we acquire 
\begin{equation}
P_{\mathrm{film}}
=
\varepsilon_0
\left[
\frac{A Q_{\mathrm{meas}}}{V}
-
B \varepsilon_{\mathrm{sub}}(T)
-
C
-
1
\right]E,
\end{equation}
where $Q_{\mathrm{meas}}$ is the charge measured, $\varepsilon_{\mathrm{sub}}$ is the dielectric constant of the substrate, $V$ is the voltage applied and $E$ is the electric field between the electrodes. Because $E$ in the CaTiO$_3$ layer lies almost
purely in-plane, we can calculate its value using $E = \frac{V}{g}$, where $g$ is the electrode spacing. The constants $A$, $B$ and $C$ are defined as {\cite{zhang2024electrocaloric}}:
\begin{equation}
A = \dfrac{1}{\varepsilon_{0}L\dfrac{K(k_2)}{K(k_2^\prime)}},
\end{equation}

\begin{equation}
B = \dfrac{\dfrac{K(k_1)}{K(k_1^\prime)} - \dfrac{K(k_2)}{K(k_2^\prime)}}{\dfrac{K(k_2)}{K(k_2^\prime)}},
\end{equation}

\begin{equation}
C = \dfrac{2\dfrac{K(k_0)}{K(k_0^\prime)} - \dfrac{K(k_1)}{K(k_1^\prime)}}{\dfrac{K(k_2)}{K(k_2^\prime)}},
\end{equation}

where $\dfrac{K(k_0)}{K(k_0^\prime)} = 0.5578$, $\dfrac{K(k_1)}{K(k_1^\prime)} = 0.5578$, and $\dfrac{K(k_2)}{K(k_2^\prime)} = 1.74\times 10^{-3}$ with our electrode geometry \cite{zhang2024electrocaloric}.

\subsection{Synchrotron x-ray diffraction}

Synchrotron x-ray diffraction measurements were performed at the I16 beamline at Diamond Light Source~\cite{collins2010diamond}, using a photon energy of 9.5~keV and a symmetric non-coplanar diffraction geometry using the 6-axis kappa diffractometer. Exposures were taken using the Pilatus3-100K area detector and these were re-mapped into reciprocal space using the msmapper program, part of the Dawn software package~\cite{basham2015data,i16_msmapper}. %

\subsection{First-principles}
The first-principles calculation has been done using the \textsc{ABINIT} software package~\cite{verstraete2025abinit} using the norm conserving~\cite{hamann1979norm} pseudo-potential with the Wu-Cohen (WC) parametrization for the exchange correlation potential~\cite{wu2006more} treating the  following orbitals as valences state: Ca $3s,3p,4s,3d$, Ti $4s,4p,4d,5s$ and O $2s, 2p$. We use an energy cut off of 40 Ha and a $8\times 8 \times 8\ \Gamma$ centered Monkhorst-Pack~\cite{monkhorst1976special} k-point mesh. The expansion of the electronic wave function by a plane wave basis has been cut off at 40 Ha. 
Our constraint relaxation has been performed using BFGS minimization algorithm by fixing two cell parameters and optimizing the remaining (both length and angle) one.
\subsection{Second-principles}

The second-principles atomistic simulations were performed using an effective interatomic potential built with the \textsc{Multibinit} package~\cite{verstraete2025abinit}, which implements the second-principles formalism introduced by Wojdeł et al.~\cite{Wojdel2013} and Escorihuela-Sayalero~et al.~\cite{escorihuela2017efficient}. This approach relies on a Taylor expansion of the potential energy surface (PES) around a reference cubic structure in terms of all relevant structural degrees of freedom. The total energy includes harmonic and anharmonic contributions associated with individual atomic displacements, macroscopic strains, and their mutual couplings, while long-range dipole–dipole interactions are treated explicitly. At the harmonic level, the expansion coefficients are obtained directly from density functional perturbation theory. At the anharmonic level, the most relevant higher-order terms are selected, and their coefficients are fitted to reproduce energies, forces, and stresses computed from density functional theory for a set of configurations designed to properly sample the PES.

The model we used is a slightly revised version  of the  CaTiO$_\mathrm{3}$ model described in Ref.~\cite{zhang2025finite} that has proven highly effective across a variety of applications~\cite{zhang2025finite,alexander2024switching}.
Only the coefficient number 22 $(Ca-O_1)_x(Ca-O_2)_z(Ca-O_3)_y$ has been slightly adjusted to better account for the tiny energy difference between $Pm$ phases at the DFT level : it is now -0.00275171 instead of -0.001574039962. This revised model is accessible at ~\cite{CTO_model}. We have also performed additional validation against DFT for the revised model: We have performed a DFT NEB calculation between the two $Pm$ phases, fitted the parameters entering the phenomenological expansion on DFT, and computed the double hysteresis loop both from DFT and second-principles approaches using Monte Carlo simulations based on the DFT-based and second-principles-based phenomenological models, respectively (see Fig.~S20 and Table~S3).

For SrTiO$_\mathrm{3}$, we have used an existing model \cite{gomez2026multimodal}. The validation of this model can be found in Ref. \cite{ULG/RVD6AL_2026}.  

All second-principles simulations have been done using \textsc{Multibinit} software.
Finite temperature simulations, using the lattice SP model, have been performed using a Hybrid molecular dynamics Monte-Carlo approach (HMC) \cite{Prokhorenko2018} using an isothermal molecular dynamics algorithm~\cite{Martyna1992} (readily implemented in \textsc{Abinit}) to generate new steps in the Markov Chain. 40 MD steps with a time propagation of 0.48 fs per step were executed between each Metropolis-Hastings Monte-Carlo evaluation. Far from transition 2000 Metropolis-Hastings Monte-Carlo evaluations revealed sufficient to generate good averages. Around the phase transitions this number was increased to 4000 evaluations. 

To relax phases of interest from ab initio and from the SP lattice model we used the Broyden-Fletcher-Goldfarb-Shanno method as implemented in \textsc{Abinit} (which has been linked to \textsc{Multibinit}). For the MD under electric field, a ther $Z^*\mathcal{E}$ was added to the SP model were $Z^*$ are the born effective charge from the cubic reference.

To extract the anisotropy coefficients \(K_1\), \(K_2\), and \(K_3\), we performed Nudged Elastic Band (NEB) calculations to find the minimum energy path connecting polarization states between the hard and easy axes. This method traces the most favorable path for polarization reorientation, revealing how the system transitions between two (meta)stable orientations under strain. For each intermediate state, along the NEB path, we extract the corresponding energy $E(\theta)$ as a function of the polarization angle $\theta$.
This set of energy values, $E(\theta)$, provides the data necessary to fit the parameters $K_1$, $K_2$, and $K_3$ in Eq. (1).

To compute the electrocaloric effect, we followed the method of Ref~\cite{graf2021unified}. Within this perturbative formalism, the adiabatic temperature change is expanded as a Taylor series in the applied field $E_\alpha$,
\begin{equation}
\Delta T(E_\alpha) \approx \frac{T^{(0)}}{C_E^{(0)}}\left[\pi^{(0)}_\alpha E_\alpha + \frac{1}{2}\pi^{(1)}_{\alpha\alpha} E_\alpha^2 + \cdots\right],
\end{equation}
where $\pi^{(0)}_\alpha$ is the spontaneous pyroelectric coefficient and $\pi^{(1)}_{\alpha\alpha}$ is the field-induced pyroelectric coefficient. When the field is applied parallel to the spontaneous polarization, $\pi^{(0)}_\alpha \neq 0$ and the linear term dominates, so that the electrocaloric response is naturally quantified in units of $\Delta T/\Delta E$. When the field is applied perpendicular to the polarization, $\pi^{(0)}_\alpha = 0$ by symmetry, as in a paraelectric or antiferroelectric state, and the leading quadratic term yields a response quantified in units of $\Delta T/\Delta E^2$.

\section*{Data Availability}
The experimental data generated in this study have been deposited  in the Yareta
database under accession code
\url{https://doi.org/10.26037/yareta:xz4nrli7tjatvj6wn5pmy7fmaq}. The
second-principles interatomic potentials for CaTiO$_3$ have been deposited  in the the ULi\`ege Open
Data Repository under accession code  \url{https://doi.org/10.58119/ULG/BITVRH}
and the second-principles interatomic potentials for SrTiO$_3$ have been deposited in the ULi\`ege Open
Data Repository under accession code  \url{https://doi.org/10.58119/ULG/RVD6AL}; together with the computational details given in the
Methods, they allow all simulation data reported in this work to be
reproduced. Any additional request should be addressed to the
corresponding authors.

\section*{Code Availability}
The data presented in this study were generated using broadly accessible first-principles and second-principles packages, as described in the Methods
section.

\newpage 
\medskip
\bibliography{reference}

\acknowledgments
The authors acknowledge Pavlo Zubko for useful discussions, Marco Lopes for technical support and Marin Alexe for provision of the closed-cycle cryostat and ferroelectric tester.
\section*{Funding Statement}
The authors acknowledge Diamond Light Source for time on Beam-
line I16 under Proposal 34148.
PhG and LB acknowledge financial support from F.R.S.-FNRS Belgium under research project TOPOTEX (Grant PDR T.0128.25) and from EU and F.R.S.-FNRS Belgium under M-ERA.NET cofund research project RESH (Grant PINT MULTI R.8012.25).
LB and PhG also acknowledge the use of the CECI supercomputer facilities funded by the F.R.S-FNRS (Grant No. 2.5020.1) and of the Tier-1 supercomputer of the Fédération Wallonie-Bruxelles funded by the Walloon Region (Grant No. 1117545). 
MH and YW acknowledge financial support from the Engineering and Physical Sciences Research Council (EPSRC) under Award no. UKRI2034.
\section*{Author Contributions Statement}

L.B., L.K., J.-M.T., M.H. and Ph.G. conceived the study and designed the methodology. L.K. deposited the films and performed their structural characterization. L.K., E.S., D.G.P., G.N. and M.H. performed the synchrotron X-ray diffraction measurements. L.K., Y.W. and M.H. performed the electrical measurements. C.T. performed additional measurements. L.B. developed the second-principles model and performed the first-principles, second-principles and phenomenological simulations. L.B., L.K., Y.W., E.S., M.H. and Ph.G. curated the data. L.B., L.K., Y.W., M.H. and Ph.G. performed the formal analysis. L.B. and L.K. prepared the figures. L.B. and Ph.G. wrote the original draft, and all authors reviewed and edited the manuscript. Ph.G., M.H. and J.-M.T. supervised the work.

\section*{Competing Interests Statement}
The authors declare no competing interests.

\end{document}